\documentclass[10pt]{iopart}
\usepackage{graphicx}
\eqnobysec

\newcommand{\beq}{\begin{equation}}
\newcommand{\eeq}{\end{equation}}
\newcommand{\beqa}{\begin{eqnarray}}
\newcommand{\eeqa}{\end{eqnarray}}

\newcommand{\abs}[1]{{\left\vert#1\right\vert}}
\renewcommand{\bar}[1]{{\overline{#1}}}
\renewcommand{\d}{{\rm d}}
\renewcommand{\e}{{\rm e}}

\newcommand{\frad}[2]{{\displaystyle{\displaystyle#1\over\displaystyle#2}}}
\newcommand{\ii}{{\rm i}}
\renewcommand{\max}{{\rm max}}
\newcommand{\mean}[1]{{\langle#1\rangle}}
\newcommand{\me}{{\e^{-\ii q}}}
\newcommand{\pe}{{\e^{\ii q}}}
\newcommand{\prob}{{\mathop{\rm Prob}}}
\renewcommand{\u}{{\bullet}}
\newcommand{\var}{{\mathop{\rm var}\,}}
\newcommand{\z}{{\circ}}
\newcommand{\C}{{\cal C}}
\newcommand{\I}{{\cal I}}
\newcommand{\J}{{\cal J}}
\newcommand{\N}{{\cal N}}
\renewcommand{\Im}{{\mathop{\rm Im}\,}}

\newcommand{\us}{u_\star}
\newcommand{\zs}{z_\star}
\newcommand{\Ss}{S_\star}

\begin{document}

\title[On the structure factor of jammed particle configurations]
{On the structure factor of jammed particle configurations on the one-dimensional lattice}

\author{Jean-Marc Luck}

\address{Universit\'e Paris-Saclay, CNRS, CEA, Institut de Physique Th\'eorique,
91191~Gif-sur-Yvette, France}

\begin{abstract}
A broad class of blocked or jammed configurations of particles on the one-dimensional lattice
can be characterized in terms of local rules
involving only the lengths of clusters of particles (occupied sites)
and of holes (empty sites).
Examples of physical relevance include
the metastable states reached by the zero-temperature dynamics of kinetically constrained spin chains,
the attractors of totally irreversible processes such as random sequential adsorption,
and arrays of Rydberg atoms in the blockade regime.
The configurational entropy of ensembles of such blocked configurations
has been investigated recently by means of an approach
inspired from the theory of stochastic renewal processes.
This approach provides a valuable alternative to the more traditional transfer-matrix formalism.
We show that the renewal approach is also an efficient tool to investigate a range of observables
in uniform ensembles of blocked configurations, besides their configurational entropy.
The main emphasis is on their structure factor and correlation function.
\end{abstract}

\ead{\mailto{jean-marc.luck@ipht.fr}}

\maketitle

\section{Introduction}
\label{intro}

Blocked or jammed configurations are met in many guises
in the statistical physics of complex systems,
including first and foremost glassy and granular materials
(see~\cite{LN,VH} for reviews).
In this work we focus our attention onto
blocked configurations of particles on the one-dimensional lattice.
Various situations of physical significance
involving such arrays of particles are recalled below.

Many examples of blocked configurations in one dimension
correspond to the attractors of some underlying zero-temperature dynamics
launched from an infinite-temperature disordered initial state.
Within this setting, these blocked configurations may be viewed as one-dimensional zero-temperature analogues
of the metastable states (also known as valleys, pure states, inherent structures, or quasi-states)
which are met in higher-dimensional or mean-field models at finite temperature
(see~\cite{bm,gl} and references therein).
A broad variety of kinetic spin models
are known to have an exponentially large number of attractors
consisting of single blocked or jammed configurations.
This includes pristine models,
such as the Ising chain with Kawasaki dynamics~\cite{cks,dsk},
disordered models, such as the Ising spin glass~\cite{dg,msj},
and a breadth of kinetically constrained spin models~\cite{pfa,pje,pse,crrs,dl1,dl2,dl3,pb,ds}.
Several lattice gas models
share the same phenomenology~\cite{pf,ef,kpri,klin,kkra}.
The zero-temperature dynamics of the above models is irreversible and strongly non-ergodic.
Some of these kinetic models can be exactly mapped onto the dynamics
of the deposition of hard objects on the one-dimensional lattice.
Prototypical examples are RSA (random sequential adsorption)
or CSA (cooperative sequential adsorption),
where particles or clusters are deposited irreversibly and sequentially
on an initially empty substrate~\cite{evans,talbot,kinetic}.
The process stops when no further object can be inserted into the system,
which is left in a blocked or jammed configuration
with a non-trivial density or coverage.

Another more recent motivation to investigate blocked configurations of particles
defined by local constraints comes from an entirely different area,
namely the physics of ultracold atoms.
Trapped Rydberg atoms appear as a promising benchmark for what concerns
quantum computation, simulation and information processing~\cite{swm}.
Their large size and strong interactions may give rise to a blockade,
preventing the excitation of Rydberg atoms
in the vicinity of an already existing one~\cite{jcz,lbr,pdl,vhb,hgs,bsk}.
In the simple setting of a one-dimensional optical lattice,
each site occupied by a Rydberg atom must have at least~$b$ empty sites on either side,
where $b\ge1$ is referred to as the blockade range~\cite{san,kld}.
To close this panorama with an example outside the realm of physics,
we mention the Riviera model,
where houses are sequentially built on an infinite array of pre-drawn plots along a beach,
with the constraint that every house should enjoy the sunlight
from at least one of the side directions.
New houses are successively introduced
until a blocked configuration is reached~\cite{hr1,hr2,hr3,epjst}.

A prominent quantity characterizing statistical ensembles of blocked configurations
is their configurational entropy $\Ss$
describing the exponential growth of the number $\N_N\sim\exp(N\Ss)$ of blocked configurations
with the system size $N$.
In many situations, the configurational entropy has been determined
either by combinatorial means~\cite{crrs,kld,d1,ld,dos1,dos2,crew}
or by the transfer-matrix approach~\cite{dl3,ds}.
We have proposed a novel method to evaluate $\Ss$
in a recent paper in collaboration with Krapivsky~\cite{us}.
This approach, inspired from the theory of renewal processes,
works whenever the rules defining blocked configurations
involve only the lengths of clusters of particles and holes.
Within this scope, which embraces most situations of physical significance,
the renewal approach is more systematic
and easier to implement than the more traditional transfer-matrix approach.

The goal of the present work is to show that the same approach
can be efficiently used to investigate a range of observables,
besides the configurational entropy.
of uniform ensembles of blocked configurations.
The setup of this paper is as follows.
Section~\ref{model} is devoted to generalities and to simple observables.
We first recall the definition of the statistical ensembles introduced in~\cite{us}
(section~\ref{themodel})
and the determination of their configurational entropy by the renewal approach
(section~\ref{confent}).
In section~\ref{locs} we use the same approach to investigate simple local observables,
such as boundary probabilities and the length distributions of clusters of particles and holes.
Section~\ref{cs} is devoted to the main subject of this work,
namely the structure factor and the correlation function of uniform ensembles.
After recalling some definitions (section~\ref{csdefs}),
we successively evaluate Fourier amplitudes (section~\ref{csamps}),
structure factors (section~\ref{cssf}),
and correlation functions (section~\ref{csc}),
describing in some detail the general properties of these quantities.
The next sections concern examples of ensembles of blocked configurations.
Explicit results are derived in section~\ref{exs}
for the three simplest statistical ensembles already considered in~\cite{us}.
We then consider the ensemble of attractors of the $k$-mer deposition model
(section~\ref{kmer})
and of arrays of Rydberg atoms with blockade range~$b$ (section~\ref{rydberg}).
Single configurations pertaining to both models
can be mapped onto each other, with the correspondence $k=b+1$.
A brief discussion is given in section~\ref{disc}.
An appendix is devoted to the Hendricks-Teller model~\cite{HT},
a simple example of a random structural model
whose structure factor can be studied by means of the renewal approach.

\section{Model and simple observables}
\label{model}

\subsection{The model}
\label{themodel}

Let us first recall the statistical model of blocked configurations introduced in~\cite{us}.
Configurations are semi-infinite sequences of particles
(or occupied sites, noted $\u$) and of holes (or empty sites, noted $\z$).
Such a configuration $\C$ is described by the lengths $i_1,j_1,i_2,j_2,\dots$
of the successive clusters of particles and holes.
It is therefore either of the form
\beq
\C_I=\underbrace{\u\cdots\u}_{i_1}
\underbrace{\z\cdots\z}_{j_1}
\underbrace{\u\cdots\u}_{i_2}
\underbrace{\z\cdots\z}_{j_2}\cdots
\label{init1}
\eeq
if the first site is occupied,
or of the form
\beq
\C_J=\underbrace{\z\cdots\z}_{j_1}
\underbrace{\u\cdots\u}_{i_1}
\underbrace{\z\cdots\z}_{j_2}
\underbrace{\u\cdots\u}_{i_2}\cdots
\label{init0}
\eeq
if the first site is empty.

Blocked configurations are defined by the following local rules:
\begin{itemize}
\item
All lengths $i_1,i_2,\dots$ of particle clusters
belong to some set $\I$ of positive integers.
\item
All lengths $j_1,j_2,\dots$ of hole clusters
belong to some set $\J$ of positive integers.
\end{itemize}
In the above, $\I$ and $\J$ are two prescribed finite or infinite subsets
of the positive integers, which entirely define the configurational model.
Hereafter we always deal with the uniform statistical ensemble
obtained by attributing equal weights to all blocked configurations $\C$
constructed as above.
We assume that the model is non-trivial,
namely that both sets~$\I$ and~$\J$ are non-empty
and at least one of them contains more than one integer.

\subsection{Configurational entropy}
\label{confent}

To set the stage, we begin by recalling the analysis of the configurational entropy.
This quantity, denoted by $\Ss$, describes the exponential growth
\beq
\N_N\sim\exp(N\Ss)
\label{nexp}
\eeq
of the number $\N_N$ of configurations $\C_N$ on a finite system of length $N$.
No overhang is allowed in the construction,
so that the rightmost cluster ends exactly at site $N$.
We also introduce the numbers $\N_{N,I}$ and $\N_{N,J}$
of finite configurations $\C_{N,I}$ and $\C_{N,J}$ beginning with a cluster of particles or of holes,
in correspondence with~(\ref{init1}) and~(\ref{init0}).

The key observation made in~\cite{us}
is the existence of a formal analogy between the present setting
and the theory of stochastic renewal processes~\cite{cox,cm,feller}
(see~\cite{glrenew,sbm} for presentations in the physics literature).
According to their usual definition,
renewal processes take place on the semi-infinite time axis $(t>0)$ and evolve from a given initial condition.
Hence considering semi-infinite configurations with prescribed initial condition
is the most appropriate framework for exploiting to the full the analogy between the present problem and renewal processes.
From a more technical viewpoint, this analogy also leads one to introduce the generating series of the numbers of configurations
\beqa
\N_I(z)=\sum_{N\ge1}\N_{N,I}z^N,\qquad\N_J(z)=\sum_{N\ge1}\N_{N,J}z^N,
\nonumber\\
\N(z)=\sum_{N\ge0}\N_Nz^N.
\eeqa
We have
\beq
\N(z)=1+\N_I(z)+\N_J(z),
\eeq
where the initial term $\N_0=1$ is conventional.
The series $\N_I(z)$ and $N_J(z)$ obey linear renewal equations of the form
\beq
\N_I(z)=I(z)(1+\N_J(z)),\qquad
\N_J(z)=J(z)(1+\N_I(z)),
\label{ninj}
\eeq
where
\beq
I(z)=\sum_{i\in\I}z^i,
\qquad
J(z)=\sum_{j\in\J}z^j
\eeq
are the series associated with the sets $\I$ and $\J$
defining the statistical ensemble.
Solving the linear equations~(\ref{ninj}), we obtain the results
\beq
\N_I(z)=\frac{I(z)(1+J(z))}{1-I(z)J(z)},\qquad
\N_J(z)=\frac{J(z)(1+I(z))}{1-I(z)J(z)}
\label{nijres}
\eeq
and
\beq
\N(z)=\frac{(1+I(z))(1+J(z))}{1-I(z)J(z)}.
\label{nres}
\eeq

As $z$ increases from 0 to 1,
the product $I(z)J(z)$ increases from $I(0)J(0)=0$
to $I(1)J(1)=\abs{\I}\abs{\J}\ge2$ (possibly infinite).
There is therefore a single value~$\zs$ of $z$ in the range $0<\zs<1$ such that
the denominator of~(\ref{nres}) vanishes, i.e.,
\beq
I(\zs)J(\zs)=1.
\eeq
The configurational entropy entering the exponential growth law~(\ref{nexp}) reads
\beq
\Ss=-\ln\zs.
\label{sent}
\eeq

Throughout the following, along the lines of~\cite{us},
we restrict this study to the rational class of models
where both generating series
\beq
I(z)=\frac{A_I(z)}{B_I(z)},\qquad
J(z)=\frac{A_J(z)}{B_J(z)}
\eeq
are rational functions, i.e., ratios of polynomials in $z$.
This class encompasses all examples considered below,
and virtually all situations of physical significance
(see~\cite{us} for more details).
The formula~(\ref{nres}) then takes the~form
\beq
\N(z)=\frac{C(z)}{D(z)},
\label{ncd}
\eeq
where $C(z)$ and $D(z)$ are polynomials in $z$.
The rational fraction~(\ref{ncd}) is assumed to be irreducible.
The degree of $D(z)$ is denoted by $\Delta$.
Its smallest zero $\zs$ is real, positive, and simple.
The degree $\Delta$ provides a natural measure of the complexity
of configurational models in the rational class.

\subsection{Local observables}
\label{locs}

Our main goal is to show that the renewal method recalled above
provides an efficient tool to evaluate
a good deal of observables in the uniform ensemble, besides its configurational entropy.
The case of the Hendricks-Teller model,
considered in~\ref{appa},
serves as a warming up for what follows.
The remainder of this section is devoted to local observables
such as boundary probabilities
and the length distribution of particle and hole clusters.
Section~\ref{cs} will be about the main subject of this work,
namely the structure factor and the correlation function.

\subsubsection*{Boundary probabilities.}

Our first observables are the boundary probabilities $W_I$ (resp.~$W_J$)
that an infinitely long configuration $\C$
begins with a cluster of particles (resp.~a cluster of holes), i.e.,
\beq
W_I=\lim_{N\to\infty}\frac{\N_{N,I}}{\N_N},\qquad
W_J=\lim_{N\to\infty}\frac{\N_{N,J}}{\N_N}.
\eeq
We have $\N_N=\N_{N,I}+\N_{N,J}$ for $N\ge1$, and so $W_I+W_J=1$, as expected.
All generating series entering~(\ref{nijres}) and~(\ref{nres}) have poles at $z=\zs$,
so that all numbers of configurations grow as $\exp(N\Ss)$.
We have therefore
\beq
W_I=\lim_{z\to\zs}\frac{\N_I(z)}{\N(z)},\qquad
W_J=\lim_{z\to\zs}\frac{\N_J(z)}{\N(z)}.
\label{ztrick}
\eeq
The formulas~(\ref{nijres}) and~(\ref{nres}) yield
\beqa
W_I=\frac{I(\zs)}{1+I(\zs)}=\frac{1}{1+J(\zs)},
\nonumber\\
W_J=\frac{J(\zs)}{1+J(\zs)}=\frac{1}{1+I(\zs)}.
\eeqa

\subsubsection*{Distributions of cluster lengths and their moments.}

We now turn to the length distributions of the clusters of particles and of holes,
i.e.,
\beq
p_{I,k}=\prob(i=k),\qquad p_{J,k}=\prob(j=k),
\eeq
where $i$ (resp.~$j$) stands for the length of any particle cluster (resp.~any hole cluster)
in an infinitely long configuration.
Consider particle clusters for definiteness.
All such clusters are interchangeable, so that we have
\beq
p_{I,k}=\prob(i_1=k)=\lim_{N\to\infty}\frac{\N_{N,I,k}}{\N_{N,I}},
\eeq
where $\N_{N,I,k}$ is the number of configurations of length $N$ beginning
with a cluster of particles of length $i_1=k$.
The corresponding generating series reads
\beq
\N_{I,k}(z)=\sum_{N\ge1}\N_{N,I,k}z^N=z^k\,\chi_I(k)(1+\N_J(z)),
\eeq
where
\beq
\chi_I(k)=\left\{
\begin{array}{ll}
1 & (k\in\I),\\
0 & (k\notin\I)
\end{array}
\right.%}
\eeq
is the indicator function of the set $\I$.
In line with~(\ref{ztrick}), we have
\beq
p_{I,k}=\lim_{z\to\zs}\frac{\N_{I,k}(z)}{\N_I(z)}.
\eeq
We thus obtain
\beq
p_{I,k}=\frac{\zs^k}{I(\zs)}\,\chi_I(k),
\eeq
and similarly
\beq
p_{J,k}=\frac{\zs^k}{J(\zs)}\,\chi_J(k).
\eeq
In particular,
the mean lengths of particle and hole clusters read
\beqa
&\mean{i}=\sum_{k\ge1}k\,p_{I,k}=\sum_{k\in\I}\frac{k\zs^k}{I(\zs)}
=\frac{\zs I'(\zs)}{I(\zs)}=\zs I'(\zs)J(\zs),
\nonumber\\
&\mean{j}=\zs I(\zs)J'(\zs).
\label{resij}
\eeqa
Here and throughout the following,
$\mean{.}$ denotes an average over the uniform ensemble of blocked configurations.
The mean distance between consecutive particle clusters
(or, equivalently, between consecutive hole clusters) is therefore
\beq
\ell=\mean{i}+\mean{j}=\zs(I'(\zs)J(\zs)+I(\zs)J'(\zs)).
\label{elldef}
\eeq
The particle density can be obtained as
\beq
\rho=\frac{\mean{i}}{\ell}=\frac{I'(\zs)J(\zs)}{I'(\zs)J(\zs)+I(\zs)J'(\zs)}.
\label{rho}
\eeq
An equivalent expression was derived by another route in~\cite{us}.
Higher moments of the cluster lengths can be determined along the same lines:
\beqa
\mean{i^p}=J(\zs)\Bigl(z\frac{\d}{\d z}\Bigr)^pI(z)\Big\vert_{z=\zs},
\nonumber\\
\mean{j^p}=I(\zs)\Bigl(z\frac{\d}{\d z}\Bigr)^pJ(z)\Big\vert_{z=\zs}.
\eeqa
We have especially
\beqa
\mean{i^2}=\zs J(\zs)(I'(\zs)+\zs I''(\zs)),
\nonumber\\
\mean{j^2}=\zs I(\zs)(J'(\zs)+\zs J''(\zs)).
\label{res2}
\eeqa

\section{General results on structure factor and correlation function}
\label{cs}

\subsection{Definitions}
\label{csdefs}

Each configuration $\C$ can be alternatively described by the occupation numbers
\beq
\eta_m=\left\{
\begin{array}{ll}
1 & (\hbox{site $m$ is occupied}),\\
0 & (\hbox{site $m$ is empty}).
\end{array}
\right.%}
\eeq
The correlation functions of the model are
$\mean{\eta_m}$, $\mean{\eta_m\eta_l}$, and so on,
where averages are taken over the uniform ensemble.
Translation invariance and clustering of correlations hold
in the bulk of the system, i.e., far from its boundary.
We have in particular
\beq
\mean{\eta_m}\to\rho,\qquad
\mean{\eta_m\eta_l}-\mean{\eta_m}\mean{\eta_l}\to C_{m-l}
\eeq
as $m$ and $l$ are simultaneously large,
the difference $m-l$ being kept fixed.
The particle density $\rho$ is given by~(\ref{rho}).
The connected pair correlation function $C_n$
is an even function of~$n$ which falls off to zero
as the distance $\abs{n}$ between both sites gets large.
It is invariant under the exchange of particles and holes
($\eta_m\longleftrightarrow1-\eta_m$).

The (random) Fourier amplitude $G_N(q)$ and intensity $S_N(q)$
associated with the first $N$ sites read
\beq
G_N(q)=\sum_{n=1}^N\e^{-\ii mq}\eta_m,\qquad
S_N(q)=\frac{\abs{G_N(q)}^2}{N}.
\label{gdefs}
\eeq
Our main interest is in the structure factor
\beq
S(q)=\lim_{N\to\infty}\mean{S_N(q)}.
\eeq
This self-averaging quantity identifies with the Fourier transform
of the correlation function $C_n$:
\beq
S(q)=\sum_{n=-\infty}^\infty C_n\e^{-\ii nq}.
\label{scorr}
\eeq

\subsection{Fourier amplitudes}
\label{csamps}

In order to determine the Fourier amplitudes,
it is advantageous to split the definition~(\ref{gdefs}),
denoting by $G_{N,I}(q)$ (resp.~$G_{N,J}(q)$)
the Fourier amplitude of a finite configuration of type $\C_{N,I}$ (resp.~$\C_{N,J}$).
These amplitudes obey the renewal equations
\beqa
G_{N,I}(q)=a(i_1,q)+\e^{-\ii qi_1}G_{N-i_1,J}(q),
\nonumber\\
G_{N,J}(q)=\e^{-\ii qj_1}G_{N-j_1,I}(q),
\label{grenew}
\eeqa
with
\beq
a(i_1,q)
=\sum_{m=1}^{i_1}\e^{-\ii mq}
=\frac{\me}{1-\me}\,\left(1-\e^{-\ii qi_1}\right).
\eeq
In~(\ref{grenew}), the amplitude denoted by $G_{N-i_1,J}(q)$
is associated with the $N-i_1$ atoms of the configuration $\C_I$
numbered from $i_1+1$ to $N$.
It is a probabilistic copy of the amplitude $G_{N-i_1,J}(q)$.
Similarly, the amplitude denoted by $G_{N-j_1,I}(q)$
is associated with the $N-j_1$ atoms of the configuration $\C_J$
numbered from $j_1+1$ to~$N$.
It is a probabilistic copy of the amplitude $G_{N-j_1,I}(q)$.

Let us introduce the generating series
\beqa
\Gamma_I(z,q)=\sum_{N\ge1}\sum_{\C_{N,I}}G_{N,I}(q)z^N,
\nonumber\\
\Gamma_J(z,q)=\sum_{N\ge1}\sum_{\C_{N,J}}G_{N,J}(q)z^N,
\nonumber\\
\Gamma(z,q)=\Gamma_I(z,q)+\Gamma_J(z,q).
\eeqa
The renewal equations~(\ref{grenew}) entail that these quantities obey
\beqa
\Gamma_I(z,q)=\gamma(z,q)+I(z\me)\Gamma_J(z,q),
\nonumber\\
\Gamma_J(z,q)=J(z\me)\Gamma_I(z,q),
\label{gammaeqs}
\eeqa
with
\beq
\gamma(z,q)=\frac{\me}{1-\me}\left(I(z)-I(z\me)\right)\left(1+\N_J(z)\right).
\label{adef}
\eeq
Solving the linear equations~(\ref{gammaeqs}), we obtain
\beqa
\Gamma_I(z,q)=\frac{\gamma(z,q)}{1-I(z\me)J(z\me)},
\nonumber\\
\Gamma_J(z,q)=\frac{J(z\me)\gamma(z,q)}{1-I(z\me)J(z\me)},
\nonumber\\
\Gamma(z,q)=\frac{(1+J(z\me))\gamma(z,q)}{1-I(z\me)J(z\me)}.
\label{gammares}
\eeqa

In the limit of an infinitely long sample,
in line with~(\ref{ztrick}),
the mean Fourier amplitude is given by
\beq
\mean{G(q)}=\lim_{z\to\zs}\frac{\Gamma(z,q)}{\N(z)}.
\eeq
Using~(\ref{nres}),~(\ref{adef}) and~(\ref{gammares}), we are left with
\beq
\mean{G(q)}=\frac{\me}{1-\me}\,\frac{1+J(\zs\me)}{1+I(\zs)}\,
\frac{I(\zs)-I(\zs\me)}{1-I(\zs\me)J(\zs\me)}.
\eeq
This result diverges at small wavevectors as
\beq
\mean{G(q)}\approx\frac{\rho}{\ii q}\qquad(q\to0),
\eeq
where the particle density $\rho$ is given by~(\ref{rho}).

\subsection{Structure factor}
\label{cssf}

The structure factor $S(q)$ encodes bulk properties of the model.
Boundary conditions are therefore irrelevant,
so that it is sufficient to consider configurations of type $\C_I$.
The simplest route to evaluate $S(q)$ consists in eliminating
amplitudes of the form $G_{N,J}(q)$
from the renewal equations~(\ref{grenew}), obtaining thus
\beq
G_{N,I}(q)=a(i_1,q)+\e^{-\ii q(i_1+j_1)}G_{N-i_1-j_1,I}(q),
\eeq
with the same convention as in~(\ref{grenew}),
and so
\beqa
G_{N,I}(q)\bar{G}_{N,I}(q)
&=a(i_1,q)\bar{a}(i_1,q)
\nonumber\\
&+a(i_1,q)\e^{\ii q(i_1+j_1)}\bar{G}_{N-i_1-j_1,I}(q)
\nonumber\\
&+\bar{a}(i_1,q)\e^{-\ii q(i_1+j_1)}G_{N-i_1-j_1,I}(q)
\nonumber\\
&+G_{N-i_1-j_1,I}(q)\bar{G}_{N-i_1-j_1,I}(q).
\label{ggfour}
\eeqa
Here and throughout the following, a bar denotes complex conjugation.

Let us introduce the generating series
\beq
\Sigma(z,q)=\sum_{N\ge1}\sum_{\C_{N,I}}G_{N,I}(q)\bar{G}_{N,I}(q)z^N.
\eeq
This quantity is given by the sum of four terms,
in correspondence with the four lines of the right-hand side of~(\ref{ggfour}), namely
\beqa
\Sigma_1(z,q)=\frac{2I(z)-I(z\pe)-I(z\me)}{2(1-\cos q)}\,\left(1+\N_J(z)\right),
\nonumber\\
\Sigma_2(z,q)=\frac{\me}{1-\me}(I(z\pe)-I(z))J(z\pe)\bar{\Gamma}_I(z,q),
\nonumber\\
\Sigma_3(z,q)=\frac{\pe}{1-\pe}(I(z\me)-I(z))J(z\me)\Gamma_I(z,q),
\nonumber\\
\Sigma_4(z,q)=I(z)J(z)\Sigma(z,q).
\label{sig123}
\eeqa
We have therefore
\beq
\Sigma(z,q)=\frac{\Sigma_1(z,q)+\Sigma_2(z,q)+\Sigma_3(z,q)}{1-I(z)J(z)}.
\label{sigres}
\eeq

In the limit of an infinitely long sample,
in line with~(\ref{ztrick}),
the structure factor is given by
\beq
S(q)=\lim_{z\to\zs}\frac{\Sigma(z,q)}{z\N_I'(z)},
\eeq
where the denominator
\beq
z\N_I'(z)=\sum_{N\ge1}N\N_{N,I}z^N
\eeq
takes care of the factor $N$ in the definition~(\ref{gdefs}) of $S_N(q)$.
This quantity has a double pole at $z=\zs$, of the form
\beq
z\N_I'(z)\approx\frac{\ell\,I(\zs)(1+J(\zs))}{(1-I(z)J(z))^2},
\label{dble}
\eeq
where the length $\ell$ is given by~(\ref{elldef}).

Some algebra using~(\ref{sig123}),~(\ref{sigres}),~(\ref{dble})
leads us to the following expression for the structure factor:
\beq
S(q)=\frac{\Phi(\pe)+\Phi(\me)}{2\ell\,(1-\cos q)},
\label{s}
\eeq
with
\beq
\Phi(\pe)=\frac{(I(\zs)-I(\zs\pe))(J(\zs)-J(\zs\pe))}{1-I(\zs\pe)J(\zs\pe)}.
\label{phi}
\eeq

The formula~(\ref{s}) is the key result of this work.
First of all,
this expression is invariant under the interchange of particles and holes, as it should be.
For all models in the rational class under consideration,
the structure factor $S(q)$ can be reduced to a rational function of~$\cos q$.
Numerous explicit examples will be given hereafter
(see~(\ref{s1}), (\ref{s2}), (\ref{s3}), (\ref{sk2}), (\ref{sk3}), (\ref{sr1}), (\ref{sr2})).
Its denominator is a polynomial in $\cos q$ whose degree is generically $\Delta-1$,
where $\Delta$ is the degree of the polynomial~$D(z)$ introduced in~(\ref{ncd}).
The structure factor has poles at the complex values $\pm q_m$ of the wavevector, such that
\beq
\e^{\ii q_m}=\frac{\zs}{z_m},
\label{qm}
\eeq
where the $z_m$ are the zeros of $D(z)$, except the smallest one $\zs$.
There are $\Delta-1$ such zeros, counted with their multiplicities.
The $z_m$ are either real, or they occur in complex conjugate pairs.
They are assumed to be ordered such that
\beq
\zs<\abs{z_1}\le\abs{z_2}\le\dots,\qquad
0<\Im q_1\le\Im q_2\le\dots
\eeq

The structure factor at zero wavevector,
\beq
S(0)=\sum_{n=-\infty}^\infty C_n,
\eeq
is nothing but the compressibility of the model.
This quantity describes the linear growth of the variance
of the total particle number,
\beq
M=\sum_{m=1}^N\eta_m,
\eeq
in a finite system of size $N$:
\beq
\var M=\mean{M^2}-\mean{M}^2\approx S(0)\,N.
\eeq
Expanding~(\ref{s}) in $q$,
and using~(\ref{resij}) and~(\ref{res2}),
we obtain
\beq
S(0)=\frac{\mean{j}^2\bigl(\mean{i^2}-\mean{i}^2\bigr)
+\mean{i}^2\bigl(\mean{j^2}-\mean{j}^2\bigr)}{\ell^3}.
%=\frac{\mean{j}^2\,\var i+\mean{i}^2\,\var j}{\ell^3}.
\label{szero}
\eeq
An equivalent formula was derived by another route in~\cite{us},
where $S(0)$ is denoted by~$c_2$.

The structure factor at wavevector $\pi$,
\beq
S(\pi)=\sum_{n=-\infty}^\infty (-1)^nC_n,
\eeq
can be referred to as the staggered compressibility.
It reads
\beq
S(\pi)=\frac{(I(\zs)-I(-\zs))(J(\zs)-J(-\zs))}{2\ell\,(1-I(-\zs)J(-\zs))}.
\eeq

The structure factor exhibits systematic extinctions
in cases where the lengths of all clusters of particles (or of holes)
are multiples of some fixed integer $K=2,3,\dots$
This phenomenon is to be expected.
Consider for definiteness the case of particle clusters.
If all their lengths are multiples of $K$,
$I(z)$ is a rational function of $z^K$,
so that the expression~(\ref{s}) of $S(q)$ vanishes whenever $\e^{\ii qK}=1$,
but $\e^{\ii q}\ne1$,
i.e., for all wavevectors of the form
\beq
q=\frac{2\pi m}{K}\qquad(m\ne0\;\hbox{mod}\;K).
\label{qext}
\eeq

In the particular case of a symmetric ensemble,
where particles and holes play similar roles,
such that $\I=\J$, $I(\zs)=J(\zs)=1$,
and $\rho=1/2$,~(\ref{s}) and~(\ref{szero}) boil down to
\beq
S(q)=\frac{1}{\ell(1-\cos q)}\,\frac{1-I(\zs\pe)I(\zs\me)}{(1+I(\zs\pe))(1+I(\zs\me))},
\eeq
\beq
S(0)=\frac{\mean{i^2}-\mean{i}^2}{4\mean{i}}.
\eeq
The above expressions bear a striking resemblance with those of the Hendricks-Teller model
(see~(\ref{sht}),~(\ref{shtzero})).

\subsection{Correlation function}
\label{csc}

The correlation function is given by (see~(\ref{scorr}))
\beq
C_n=\int_0^{2\pi}\frac{\d q}{2\pi}\,\e^{\ii nq}S(q).
\label{cint}
\eeq
Setting $y=\e^{\ii q}$ and using~(\ref{s}), this reads
\beq
C_n=-\frac{1}{\ell}\oint\frac{\d y}{2\pi\ii}\,\frac{y^n}{(y-1)^2}\bigl(\Phi(y)+\Phi(1/y)\bigr).
\label{ccint}
\eeq
The integration contour in~(\ref{ccint}) is the unit circle,
along which the integrand is regular.
The differential element $\d y/(y-1)^2$ is invariant if $y$ is changed into $1/y$.
This ensures the expected symmetry $C_n=C_{-n}$.

Henceforth we consider $n\ge0$ for definiteness.
The integral in~(\ref{ccint}) receives contributions
from the poles of $\Phi(1/y)$ at $y_m=\e^{\ii q_m}=\zs/z_m$ (see~(\ref{qm})).

Let us assume for a while that all the zeros $z_m$ are simple and that $\Phi(\infty)$ is finite.
In this situation, the correlation function is a finite sum of decaying exponentials:
\beq
C_n=\sum_{m=1}^{\Delta-1}A_m\Bigl(\frac{\zs}{z_m}\Bigr)^n\qquad(n\ge0),
\label{csum}
\eeq
with amplitudes
\beq
A_m=\frac{\zs}{\ell}
\,\frac{2-I(\zs)J(z_m)-I(z_m)J(\zs)}{(\zs-z_m)^2(I'(z_m)J(z_m)+I(z_m)J'(z_m))}.
\eeq
The correlation therefore decays exponentially with distance, as
\beq
C_n\sim\e^{-n/\xi},
\label{cnexp}
\eeq
where the correlation length reads
\beq
\xi=\frac{1}{\ln\frad{\abs{z_1}}{\zs}}=\frac{1}{\Im q_1}.
\label{xidef}
\eeq

Let us now relax the two assumptions above.
If the zero $z_m$ is not simple, but has multiplicity $m\ge2$,
the corresponding amplitude $A_m$ in~(\ref{csum})
is replaced by a polynomial in $n$ of degree at most $m-1$.
If $\Phi(\infty)$ is not finite,
then $\Phi(y)$ grows as~$y^\mu$ at large $y$, where $\mu$ is some positive integer.
As a consequence, the formula~(\ref{csum}) for~$C_n$
might be modified for the first few values of the distance $n$ ($n=0,\dots,\mu-1$).
In all cases, the correlation length is given by~(\ref{xidef}).

The correlation function admits an appealing alternative expression,
which holds in full generality.
Consider the function $\Phi(\pe)$ defined in~(\ref{phi}).
This function equals unity in the limit $\pe\to0$,
whereas it vanishes for $q\to0$ as
\beq
\Phi(\pe)\approx-\ii qR,
\eeq
with
\beq
R=\frac{\zs I'(\zs)J'(\zs)}{I'(\zs)J(\zs)+I(\zs)J'(\zs)}
=\frac{\mean{i}\mean{j}}{\ell}=\rho(1-\rho)\ell.
\label{rdef}
\eeq
Moreover, the function $\Phi(\pe)$ is regular at least for $\abs{\pe}<1$.
Its poles are indeed situated at $\e^{-\ii q_m}$ (see~(\ref{qm})).
It can therefore be expanded~as
\beq
\Phi(\pe)=\sum_{a=1}^\infty r_a(1-\e^{\ii aq}).
\eeq
The amplitudes $r_a$ are real, but not positive in general.
They obey the sum rules
\beq
\sum_{a=1}^\infty r_a=1,\qquad
\sum_{a=1}^\infty ar_a=R.
\label{sumr}
\eeq
Moreover, the rational expression~(\ref{phi}) of $\Phi(\pe)$
implies that the amplitudes $r_a$ obey a linear recursion of the form
\beq
\lambda_0r_a+\lambda_1r_{a+1}+\lambda_2r_{a+2}+\cdots=0,
\label{rrec}
\eeq
except for finitely many initial values of $a$.
The number of terms involved in this recursion is generically $\Delta$,
the degree of the polynomial $D(z)$ introduced in~(\ref{ncd}).
The coefficients $\lambda_0,\,\lambda_1,\,\lambda_2\,\dots$
are constants dictated by the model.

In terms of the amplitudes $r_a$, the formula~(\ref{s}) for the structure factor reads
\beq
S(q)=\frac{1}{\ell}\sum_{a=1}^\infty r_a\,\frac{1-\cos aq}{1-\cos q}.
\eeq
Inserting this expression into~(\ref{cint}), we obtain
\beq
C_n=\frac{1}{\ell}\sum_{a=1}^\infty r_aI_{a,n},
\eeq
with
\beq
I_{a,n}=\int_0^{2\pi}\frac{\d q}{2\pi}\,\e^{\ii nq}\,\frac{1-\cos aq}{1-\cos q}
=\left\{
\begin{array}{cl}
a-\abs{n} & (a>\abs{n}),\\
0 & (a\le\abs{n}).
\end{array}
\right.%}
\eeq
Let us henceforth consider $n\ge0$.
The correlation function has the following expression
in terms of the amplitudes $r_a$:
\beq
C_n=\frac{1}{\ell}\sum_{a=n+1}^\infty(a-n)r_a.
\label{cr}
\eeq
Using~(\ref{rdef}) and~(\ref{sumr}), this formula can be recast as
\beq
C_n=\rho(1-\rho)-\frac{1}{\ell}\biggl(n-\sum_{a=1}^{n-1}(n-a)r_a\biggr).
\label{crfini}
\eeq
The correlation function therefore obeys the three-term recursion
\beq
2C_n-C_{n-1}-C_{n+1}=-\frac{r_n}{\ell}\qquad(n\ge1).
\eeq
As a consequence, the $C_n$ also obey the linear recursion relation~(\ref{rrec}), i.e.,
\beq
\lambda_0C_n+\lambda_1C_{n+1}+\lambda_2C_{n+2}+\cdots=0,
\label{crec}
\eeq
except for finitely many initial values of $n$,
and $C_n$ and $r_n$ share the same exponential decay~(\ref{cnexp}).

The expression~(\ref{crfini}) yields in particular
\beq
C_0=\rho(1-\rho),\qquad
C_1=\rho(1-\rho)-\frac{1}{\ell}.
\label{c0c1}
\eeq
These two results are expected.
The first one is obvious from the definition of $C_0$.
The second one expresses that the densities of $\u\z$ and $\z\u$ cluster interfaces
have the expected value, namely
\beq
\lim_{m\to\infty}\mean{\eta_m(1-\eta_{m+1})}
=\rho-\rho^2-C_1=\frac{1}{\ell}.
\eeq
At larger distances ($n\ge2$), the correlation function is not universal,
in the sense that it depends on the individual amplitudes $r_a$.
We have
\beq
C_2=\rho(1-\rho)-\frac{2-r_1}{\ell},\qquad
C_3=\rho(1-\rho)-\frac{3-2r_1-r_2}{\ell},
\eeq
and so on.

\section{Three simple examples}
\label{exs}

In this section we illustrate the above general results on the structure factor $S(q)$
and on the correlation function $C_n$
by considering three simple examples of statistical ensembles,
already studied in~\cite{us},
which virtually exhaust all cases with $\Delta=1$ or $\Delta=2$.

\subsection{Flat ensemble}
\label{flat}

The simplest statistical ensemble of all is the flat one,
where all cluster lengths are permitted.
This corresponds to $\I=\J=\{1,2,3,\dots\}$, hence
\beq
I(z)=J(z)=\frac{z}{1-z},
\eeq
so that
\beq
D(z)=1-2z,
\eeq
and so $\Delta=1$.
We have
\beq
\zs=\frac{1}{2},\qquad\rho=\frac{1}{2},\qquad\ell=4.
\eeq
Equation~(\ref{s}) reads
\beq
S(q)=\frac{1}{4},
\label{s1}
\eeq
and so
\beq
C_n=\frac{\delta_{n0}}{4},
\eeq
where $\delta_{ab}$ is the Kronecker symbol.

The absence of correlations between the occupancies of different sites
reflects the property that each site of the lattice is independently either occupied or empty.
Concomitantly, the sum entering~(\ref{csum}) is empty for $\Delta=1$.

\subsection{Isolated empty sites}
\label{isola}

Our second example is defined by the condition that empty sites (i.e., holes) are isolated.
The ensemble where occupied sites (i.e., particles) are isolated is related
to the present one by exchanging the roles of particles and holes.
Both models share the same structure factor and correlation function.
These ensembles have been considered in several contexts,
including the attractors of repulsion processes~\cite{rep}
and packings of disks in narrow channels~\cite{moo1,moo2}.

This example corresponds to $\I=\{1,2,3,\dots\}$ and $\J=\{1\}$, hence
\beq
I(z)=\frac{z}{1-z},\qquad J(z)=z,
\eeq
so that
\beq
D(z)=1-z-z^2,
\eeq
and so $\Delta=2$.
We have
\beqa
\zs&=\frac{\sqrt{5}-1}{2}\approx0.618033,
\label{gold}
\\
\rho&=\frac{5+\sqrt{5}}{10}\approx0.723606,\qquad
\ell=\frac{5+\sqrt{5}}{2}\approx3.618033.
\eeqa
Equation~(\ref{s}) reads
\beq
S(q)=\frac{1}{\sqrt{5}\,(3+2\cos q)},
\label{s2}
\eeq
and so
\beq
C_n=\frac{1}{5}\left(-\frac{3-\sqrt{5}}{2}\right)^n\qquad(n\ge0).
\eeq
The correlation function is a pure decaying exponential,
in agreement with~(\ref{csum}) for $\Delta=2$.

\subsection{Even particle clusters}
\label{even}

Our third example is defined by the condition that clusters of particles have even lengths.
This corresponds to
$\I=\{2,4,6,\dots\}$ and $\J=\{1,2,3,\dots\}$, i.e.,
\beq
I(z)=\frac{z^2}{1-z^2},\qquad J(z)=\frac{z}{1-z}.
\eeq
Here again,
\beq
D(z)=1-z-z^2
\eeq
and $\Delta=2$, so that $\zs$ is given by~(\ref{gold}).
We have
\beq
\rho=\frac{5-\sqrt{5}}{5}\approx0.552786,\qquad
\ell=\frac{5+3\sqrt{5}}{2}\approx5.854101.
\eeq
Equation~(\ref{s}) reads
\beq
S(q)=\frac{2\,(1+\cos q)}{\sqrt{5}\,(3+2\cos q)}.
\label{s3}
\eeq
The structure factor vanishes for $q=\pi$.
This is an extinction of the form~(\ref{qext}) with $K=2$.
The correlation function reads
\beq
C_n=\frac{\delta_{n0}}{\sqrt{5}}-\frac{1}{5}\left(-\frac{3-\sqrt{5}}{2}\right)^n\qquad(n\ge0).
\eeq
The expression for $n=0$ is modified with respect to the decaying exponential~(\ref{csum}).
This is an instance of the situation described below~(\ref{xidef}), with $\mu=1$.

\begin{figure}
\begin{center}
\includegraphics[angle=0,width=0.7\linewidth,clip=true]{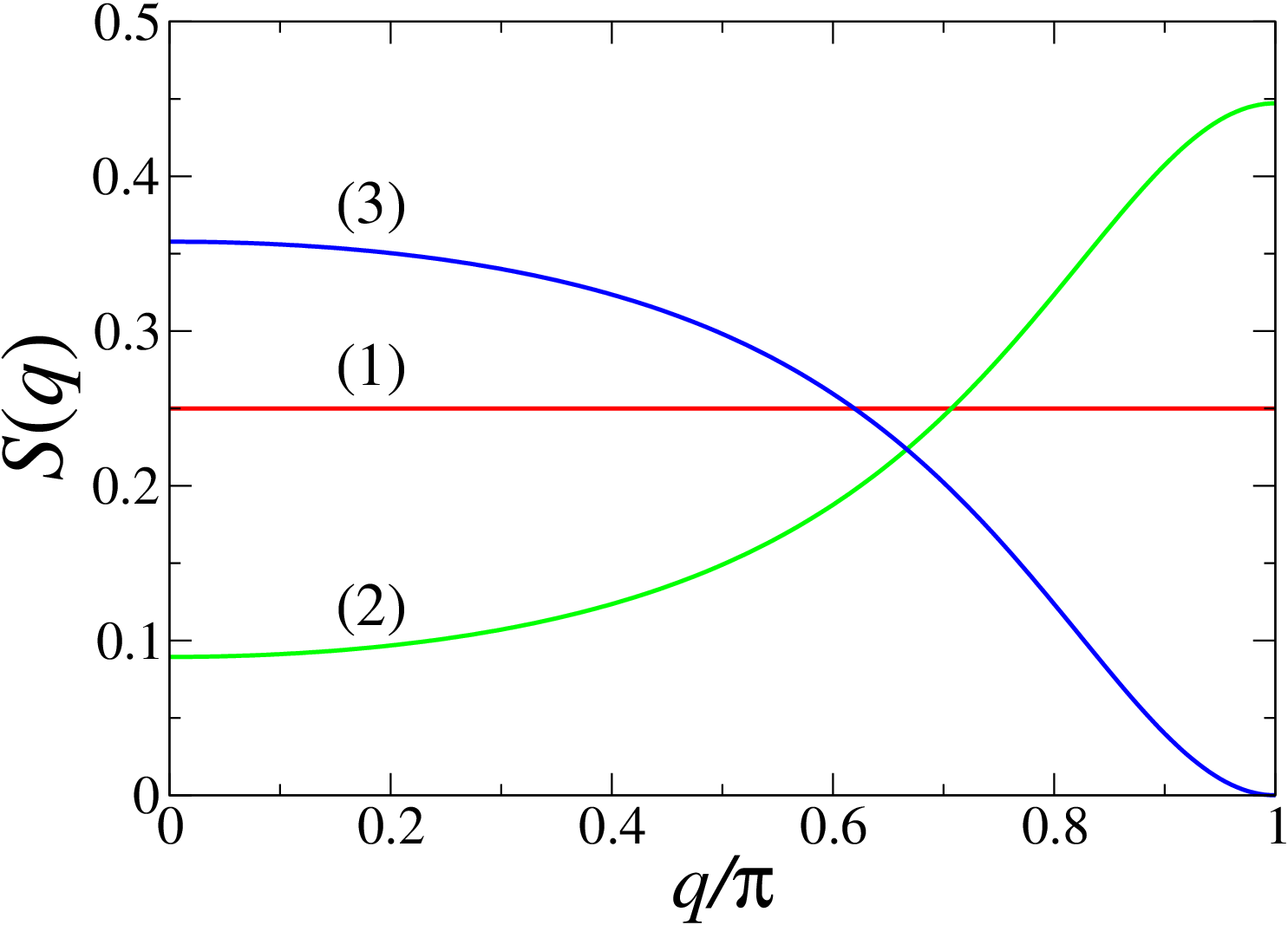}
\caption{\small
Structure factor $S(q)$
of the three ensembles investigated in section~\ref{exs},
plotted against $q/\pi$.
Red curve (1): flat ensemble (see~(\ref{s1})).
Green curve (2): isolated empty sites (see~(\ref{s2})).
Blue curve (3): even particle clusters (see~(\ref{s3})).}
\label{examples}
\end{center}
\end{figure}

Figure~\ref{examples} shows plots of the structure factor $S(q)$
against $q/\pi$ for the three ensembles investigated in this section.

\section{The $k$-mer deposition model}
\label{kmer}

The sequential deposition of $k$-mers is a fully irreversible process
where $k$-mers (clusters of~$k$ particles) are deposited at random positions
on an initially empty one-dimensional lattice.
The process stops when the system reaches a blocked (or jammed) configuration
where no further $k$-mer can be inserted any more~\cite{ghh,beg,bonnier}.
The integer $k\ge2$ is the only parameter of the model.
The sequential deposition of dimers ($k=2$) is equivalent to a dimerization model
investigated long ago by Flory~\cite{flory}.

The statistical ensemble of blocked configurations of $k$-mers
has been studied by various approaches~\cite{dos1,dos2,kld,us,crew}.
In~\cite{us} it has been shown that this ensemble
can be described in terms of independent clusters,
and therefore studied within the present renewal formalism,
with $\I=\{k,2k,3k,\dots\}$ and $\J=\{1,2,\dots,k-1\}$, so that
\beq
I(z)=\frac{z^k}{1-z^k},\qquad J(z)=\frac{z-z^k}{1-z}.
\eeq
The polynomial $D(z)$ can be reduced to
\beq
D(z)=1-(z^k+z^{k+1}+\cdots+z^{2k-1}),
\label{dkmer}
\eeq
and so
\beq
\Delta=2k-1.
\label{deltak}
\eeq
For all values of the integer $k$,
the structure factor is expected to exhibit systematic extinctions
at wavevectors that are multiples of $2\pi/k$ (see~(\ref{qext})).

\subsection{The first few values of $k$}

It is interesting to start by considering the first few values of $k$,
where closed-form formulas can be derived.

\subsubsection*{Dimers ($k=2$).}

This case is the simplest of all.
We have
\beq
D(z)=1-z^2-z^3,
\label{dk2}
\eeq
and so $\Delta=3$.
We thus recover that
\beq
\zs=\frac{1}{6}\bigl((100+12\sqrt{69})^{1/3}+(100-12\sqrt{69})^{1/3}-2\bigr)\approx0.754877
\label{zk2}
\eeq
is the reciprocal of the so-called plastic number~\cite{us},
whereas
\beq
\rho=\frac{2(7+2\zs+3\zs^2)}{23}\approx0.822991.
\eeq

Even in this simple case,~(\ref{s}) yields a rather lengthy expression for the structure factor.
Repeated use of $D(\zs)=0$
reduces this expression to a minimal form
where the highest power of $\zs$ is $\Delta-1$ (here, 2).
We shall do this reduction throughout the following.
We thus obtain
\beq
S(q)=\frac{2\zs(1-\zs)(1+\cos q)}{A_2(\cos q)},
\label{sk2}
\eeq
with
\beqa
A_2(\cos q)&=&3-\zs^2-2(4-5\zs-5\zs^2)\cos q
\nonumber\\
&-&4(3-2\zs-4\zs^2)\cos^2q.
\label{a2}
\eeqa
The extinction at $q=\pi$ is clearly visible on~(\ref{sk2}).

\subsubsection*{Trimers ($k=3$).}

In this case we have $\Delta=5$, but $D(z)$ factors as
\beq
D(z)=(1+z^2)(1-z^2-z^3),
\label{dk3}
\eeq
so that $\zs$ is again given by~(\ref{zk2}),
whereas
\beq
\rho=\frac{3(26+7\zs-2\zs^2)}{115}\approx0.786377.
\eeq
Equation~(\ref{s}) yields
\beq
S(q)=\frac{(1+2(1-\zs)\cos q)(1+2\cos q)^2}{A_3(\cos q)},
\label{sk3}
\eeq
with
\beqa
A_3(\cos q)&=&-(1-2\zs)(2+3\zs)+2(9+\zs-13\zs^2)\cos q
\nonumber\\
&+&4(4-3\zs)(2+3\zs)\cos^2q-8(3-6\zs-10\zs^2)\cos^3q
\nonumber\\
&-&16(1-2\zs)(2+3\zs)\cos^4q.
\label{a3}
\eeqa
The extinction at $q=2\pi/3$ is clearly visible on~(\ref{sk3}).

\begin{figure}
\begin{center}
\includegraphics[angle=0,width=0.7\linewidth,clip=true]{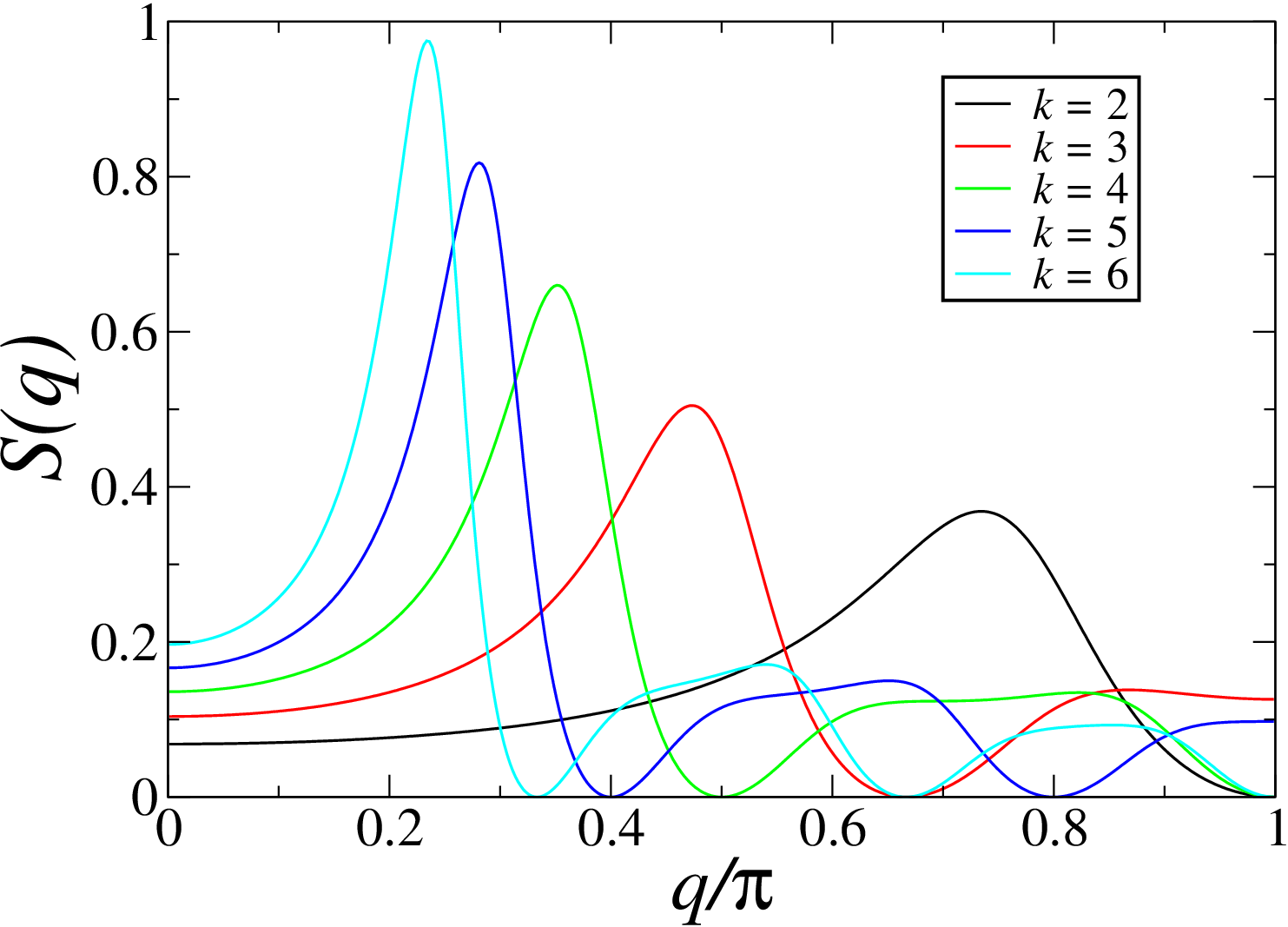}
\caption{\small
Structure factor $S(q)$ of the $k$-mer deposition model,
plotted against $q/\pi$, for $k$ ranging from 2 to 6 (see legend).}
\label{kmers}
\end{center}
\end{figure}

The formulas~(\ref{sk2}) and~(\ref{sk3})
show that the complexity of the analytical expression of $S(q)$
increases fast with the integer $k$.
The structure factor indeed exhibits more and more detailed structures as $k$ increases,
as testified by figure~\ref{kmers}.

\subsection{Scaling behavior at large $k$}
\label{klarge}

Static observables of the $k$-mer deposition model
have been shown in~\cite{us} to exhibit an unusual behavior at large $k$,
where logarithmic corrections to scaling are ubiquitous.
These corrections to scaling already show up in $\zs$.
Setting
\beq
\zs=\exp\left(-\frac{\us}{k}\right),
\label{udef}
\eeq
we find that $\us$ obeys the transcendental equation
\beq
\us\e^{\us}\approx k.
\label{usw}
\eeq
This equation holds up to corrections of relative order $1/k$,
which will be consistently neglected throughout the following.
The solution to~(\ref{usw}) reads
\beq
\us\approx W(k),
\eeq
in terms of the Lambert $W$ function.
To leading order, this reads
\beq
\us\approx\ln k.
\eeq
More precisely, setting
\beq
\lambda=\ln k,\qquad\mu=\ln\lambda=\ln\ln k,
\eeq
we obtain the full asymptotic series
\beq
\us=\lambda-\mu+\frac{\mu}{\lambda}+\frac{\mu(\mu-2)}{2\lambda^2}
+\frac{\mu(2\mu^2-9\mu+6)}{6\lambda^3}+\cdots
\eeq
The main outcomes of the scaling analysis performed in~\cite{us}
which are relevant for the present purpose are as follows:
\beq
\rho\approx\frac{\us}{\us+1},\qquad
\ell\approx\frac{\us+1}{\us}\,k.
\label{rhoell}
\eeq

The scaling analysis recalled above can be extended to the structure factor
all over its scaling regime where $q$ scales as the inverse of $k$.
Introducing the rescaled wavevector
\beq
Q=kq,
\eeq
we have the estimates
\beqa
I(\zs)\approx\frac{\us}{k},\qquad
I(\zs\e^{\pm\ii q})\approx\frac{\us\,\e^{\pm\ii Q}}{k},
\nonumber\\
J(\zs)\approx\frac{k}{\us},\qquad
J(\zs\e^{\pm\ii q})\approx\frac{k}{\us\mp\ii Q},
\eeqa
so that~(\ref{s}) evaluates to
\beq
S(q)\approx k\,\frac{\us}{\us+1}
\,\frac{2(1-\cos Q)}{2\us^2(1-\cos Q)+2\us Q\sin Q+Q^2}.
\label{sksca}
\eeq
The second factor of this expression is equal to the particle density $\rho$
(see~(\ref{rhoell})).
The third factor mainly depends on the rescaled wavevector~$Q$.
The presence of powers of $\us$
however generates logarithmic corrections to this main scaling behavior.
We have in particular
\beq
S(0)\approx\frac{k\us}{(\us+1)^3}.
\label{sk0}
\eeq
An equivalent formula was derived by another route in~\cite{us}.
The expression
\beq
C_0=\rho(1-\rho)\approx\frac{\us}{(\us+1)^2}
\eeq
translates to the non-trivial identity
\beq
\int_{-\infty}^\infty\d Q\,\frac{1-\cos Q}{2\us^2(1-\cos Q)+2\us Q\sin Q+Q^2}
=\frac{\pi}{\us+1}.
\eeq

The scaling form~(\ref{sksca}),
with its fast decay in $1/Q^2$,
corroborates the most salient feature of figure~\ref{kmers},
namely that most of the intensity concentrates onto smaller and smaller values of $q$
as $k$ is increased.

\begin{figure}
\begin{center}
\includegraphics[angle=0,width=0.7\linewidth,clip=true]{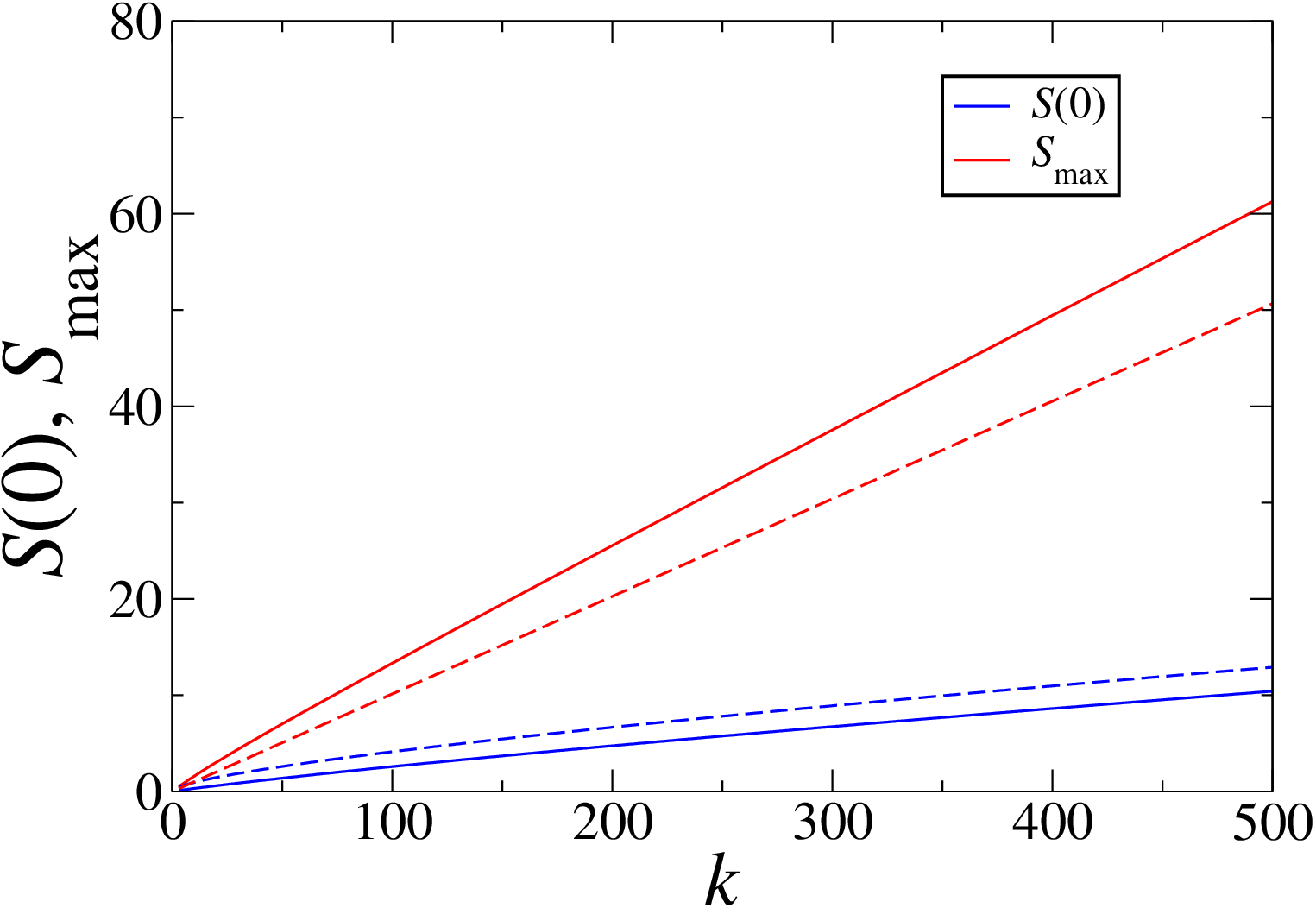}
\caption{\small
Plots of $S(0)$ and $S_\max$ against $k$, for the $k$-mer deposition model.
Full blue curve: exact values of $S(0)$ for finite $k$ (see~(\ref{szero})).
Dashed blue curve: scaling prediction (see~(\ref{udef}),~(\ref{sk0})).
Full red curve: numerically exact maximal values of $S(q)$ for finite $k$.
Dashed red curve: leading linear growth in $k/\pi^2$ (see~(\ref{sasy})).}
\label{ksca}
\end{center}
\end{figure}

Let us focus our attention onto the leading peak of the structure factor,
located slightly before the first extinction at $q=2\pi/k$,
whose height grows rather fast with~$k$.
The scaling behavior of the position $Q_\max=kq_\max$ of this peak
and of its height~$S_\max$ can be extracted from the scaling result~(\ref{sksca}).
The peak position obeys the transcendental equation
\beq
\us=\frac{Q_\max(2(1-\cos Q_\max)-Q_\max\sin Q_\max)}{2(1-\cos Q_\max)(Q_\max-\sin Q_\max)}.
\eeq
Some algebra yields the following asymptotic series
in inverse powers of $\us$:
\beqa
Q_\max&=&2\pi\left(1-\frac{1}{\us}+\frac{1}{\us^2}
+\frac{\pi^2-3}{3\us^3}+\cdots\right),
\\
S_\max&=&\frac{k}{\pi^2}\left(1+\frac{1}{\us}
-\frac{2\pi^2}{3\us^3}+\cdots\right).
\label{sasy}
\eeqa
The peak height $S_\max$ therefore grows linearly with $k$,
i.e., faster than $S(0)\approx k/(\ln k)^2$ (see~(\ref{sk0}))
by a factor growing as $(\ln k)^2$.
This is illustrated in figure~\ref{ksca},
showing plots of $S(0)$ and $S_\max$ against~$k$.
Dashed lines with corresponding colors
show appropriate approximations (see caption).

The peak position $Q_\max$ is very near the first pair of complex poles of the structure factor,
i.e., $Q_1=kq_1$ and $\bar{Q}_1=k\bar{q}_1$, where
\beq
Q_1=2\pi\left(1-\frac{1}{\us}+\frac{1+\ii\pi}{\us^2}
-\frac{4\pi^2+3+9\ii\pi}{3\us^3}+\cdots\right).
\eeq
Taking the imaginary part of this expression to leading order in $1/\us$
and using~(\ref{xidef}),
we obtain that the correlation length scales as
\beq
\xi\approx\frac{k\us^2}{2\pi^2}\approx\frac{k(\ln k)^2}{2\pi^2}.
\label{xisca}
\eeq

\section{Arrays of Rydberg atoms}
\label{rydberg}

This section is devoted to blocked configurations of arrays of Rydberg atoms
on a one-dimensional optical lattice.
We consider the simple setting recalled in section~\ref{intro},
where each lattice site occupied by a Rydberg atom
must have at least~$b$ empty sites on either side.
The integer $b\ge1$, referred to as the blockade range,
is the only parameter of the model.

Our aim is to investigate the uniform ensemble of all blocked configurations,
where no further atom can be inserted into the system.
Along the lines of earlier works~\cite{kld,crew,us},
blocked configurations consist of isolated occupied sites (the Rydberg atoms)
separated by clusters of holes whose length is at least $b$,
in order to obey the blockade constraint, and at most~$2b$,
since an extra Rydberg atom could be inserted in the middle of an empty range of size $2b+1$.

Within the present formalism,
clusters of particles and of holes respectively correspond to
$\I=\{1\}$ and $\J=\{b,b+1,\dots,2b\}$, and so
\beq
I(z)=z,\qquad J(z)=\frac{z^b(1-z^{b+1})}{1-z}.
\label{rsers}
\eeq
The polynomial $D(z)$ can be reduced to
\beq
D(z)=1-(z^{b+1}+z^{b+2}+\cdots+z^{2b+1}).
\label{drydberg}
\eeq
This expression coincides with~(\ref{dkmer}),
with the identification
\beq
k=b+1.
\label{kb}
\eeq
We have in particular
\beq
\Delta=2b+1.
\label{deltab}
\eeq

There is indeed an equivalence, at the level of single configurations,
between blocked configurations of $k$-mers and of Rydberg atoms.
A Rydberg atom followed by~$b$ empty sites (to its right) can be mapped onto a $k$-mer,
where $b$ and $k$ are related by~(\ref{kb}).
The mapping between configurations of $k$-mers and of Rydberg atoms
is however not unique~\cite{crew,us}.
Many observables pertaining to both models are simply related to each other.
For instance, the density of the $k$-mer deposition model
is $k$ times larger than the corresponding density of Rydberg atoms.

As a consequence of the above correspondence,
the structure factors of $k$-mers and Rydberg atoms with $b=k-1$ share the same denominator.
The latter is indeed a polynomial in $\cos q$ that is essentially dictated by $D(z)$.
Both structure factors have different numerators,
if only because there are no extinctions in the case of Rydberg atoms.

\subsection{The first few values of $b$}

In is again interesting to first look at a few smallest values of the blockade range~$b$.

\subsubsection*{$b=1$.}

This case is the simplest of all.
It corresponds to $k=2$,
so that~(\ref{dk2}) and~(\ref{zk2}) still hold,
whereas
\beq
\rho=\frac{7+2\zs+3\zs^2}{23}\approx0.411495.
\eeq
Equation~(\ref{s}) yields
\beq
S(q)=\frac{1}{B_1(\cos q)},
\label{sr1}
\eeq
with
\beqa
B_1(\cos q)&=&5+7\zs+5\zs^2+2(7+8\zs+2\zs^2)\cos q
\nonumber\\
&+&4(2+3\zs)\cos^2q.
\eeqa
We have (see~(\ref{a2}))
\beq
B_1(\cos q)=(2+3\zs+2\zs^2)A_2(\cos q).
\eeq
It is indeed expected from the correspondence between $k$-mers and Rydberg atoms
that the denominators of~(\ref{sk2}) and~(\ref{sr1}) should coincide,
up to some $q$-independent multiplicative constant.

\subsubsection*{$b=2$.}

This case corresponds to $k=3$,
so that~(\ref{zk2}) and~(\ref{dk3}) still hold,
whereas
\beq
\rho=\frac{26+7\zs-2\zs^2}{115}\approx0.262125.
\eeq
Equation~(\ref{s}) yields
\beq
S(q)=\frac{(1+\zs)^2+2\zs\cos q}{B_2(\cos q)},
\label{sr2}
\eeq
with
\beqa
B_2(\cos q)&=&5+3\zs-\zs^2-2(3-6\zs-10\zs^2)\cos q
\nonumber\\
&+&4(5+13\zs+14\zs^2)\cos^2q+8(13+10\zs+3\zs^2)\cos^3q
\nonumber\\
&+&16(5+3\zs-\zs^2)\cos^4q.
\eeqa
We have (see~(\ref{a3}))
\beq
B_2(\cos q)=(1+\zs)^2A_3(\cos q).
\eeq
A relationship of this kind was again expected from the correspondence
between $k$-mers and Rydberg atoms.

\begin{figure}
\begin{center}
\includegraphics[angle=0,width=0.7\linewidth,clip=true]{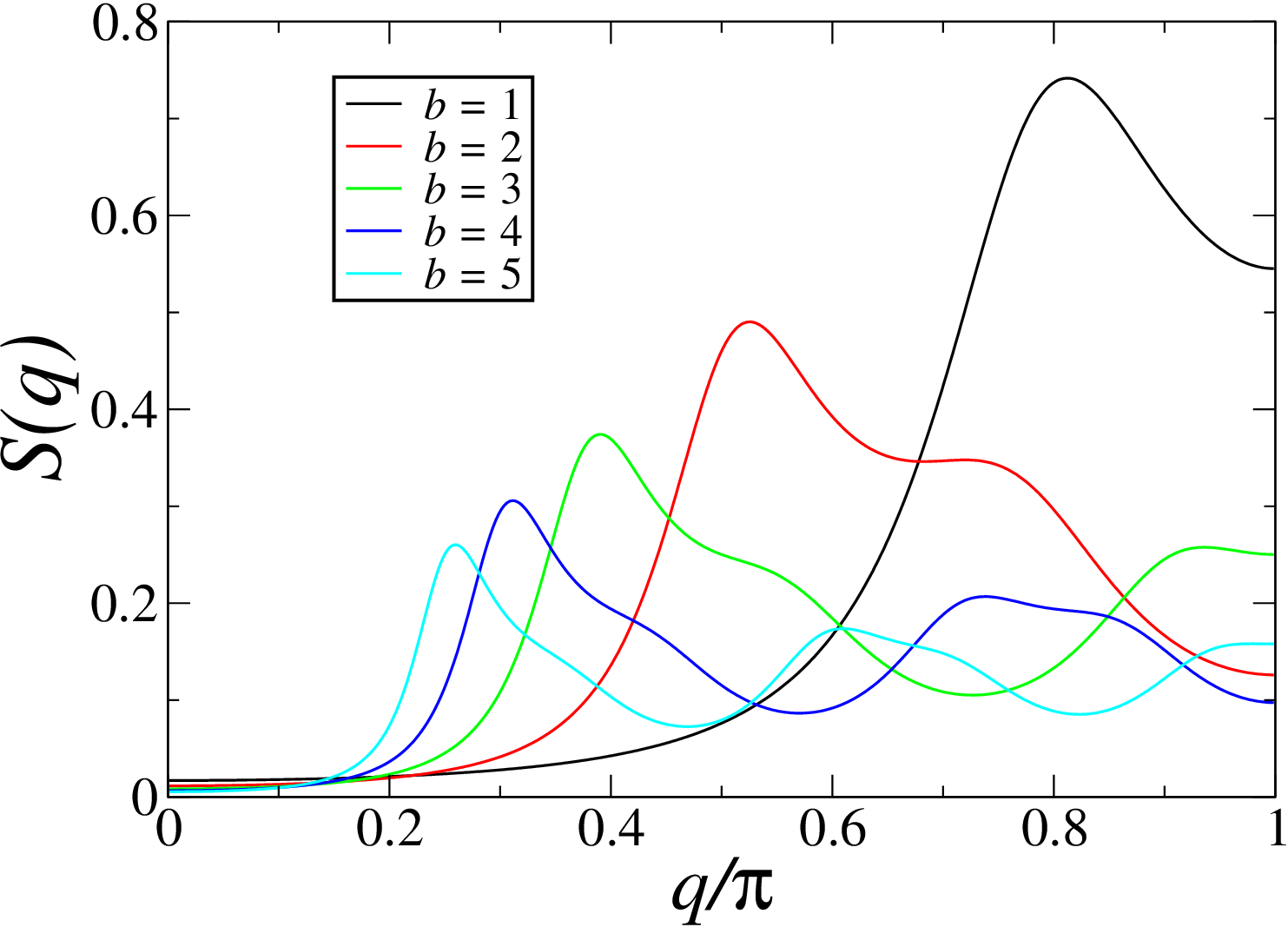}
\caption{\small
Structure factor $S(q)$ of the blocked arrays of Rydberg atoms,
plotted against $q/\pi$, for blockade ranges $b$ running from 1 to 5 (see legend).}
\label{ryd}
\end{center}
\end{figure}

Equations~(\ref{sr1}) and~(\ref{sr2})
demonstrate that the complexity of the structure factor
increases fast with the blockade range $b$.
Concomitantly, $S(q)$ exhibits more and more detailed structures as $b$ increases
(see figure~\ref{ryd}).

\subsection{Scaling behavior at large $b$}

The scaling analysis of the $k$-mer deposition model at large $k$
summarized in section~\ref{klarge} applies, mutatis mutandis,
to the blocked configurations of Rydberg atoms
in the regime where the blockade range $b$ becomes large~\cite{us}.
Here, too, logarithmic corrections to scaling are ubiquitous.
Setting
\beq
\zs=\exp\left(-\frac{\us}{b}\right),
\label{ubdef}
\eeq
we find that $\us$ again obeys the transcendental equation~(\ref{usw}),
and we have
\beq
\rho\approx\frac{\us}{b(\us+1)},\qquad
\ell\approx\frac{\us+1}{\us}\,b.
\label{brhoell}
\eeq

The scaling analysis again extends to the structure factor
all over the regime where~$q$ scales as the inverse of the blockade range $b$.
Introducing the rescaled wavevector
\beq
Q=bq,
\eeq
and skipping every detail, we find that~(\ref{s}) translates to
\beq
S(q)\approx\frac{\us}{b(\us+1)}
\,\frac{Q^2}{2\us^2(1-\cos Q)+2\us Q\sin Q+Q^2}.
\label{sbsca}
\eeq
The first factor is equal to the density $\rho$ of Rydberg atoms (see~(\ref{brhoell})).
The second factor mainly depends on the rescaled wavevector~$Q$.
The denominators of~(\ref{sksca}) and~(\ref{sbsca}) coincide,
as expected from the correspondence recalled above
between $k$-mers and Rydberg atoms.
The numerators are however different.
The presence of powers of $\us$ in~(\ref{sbsca})
again generates logarithmic corrections of various kinds.
We have in particular
\beq
S(0)\approx\frac{\us}{b(\us+1)^3}.
\label{sb0}
\eeq
The correlation length is again described by the scaling form~(\ref{xisca}),
up to the replacement of $k$ by $b$, i.e.,
\beq
\xi\approx\frac{b\us^2}{2\pi^2}\approx\frac{b(\ln b)^2}{2\pi^2}.
\eeq

\begin{figure}
\begin{center}
\includegraphics[angle=0,width=0.7\linewidth,clip=true]{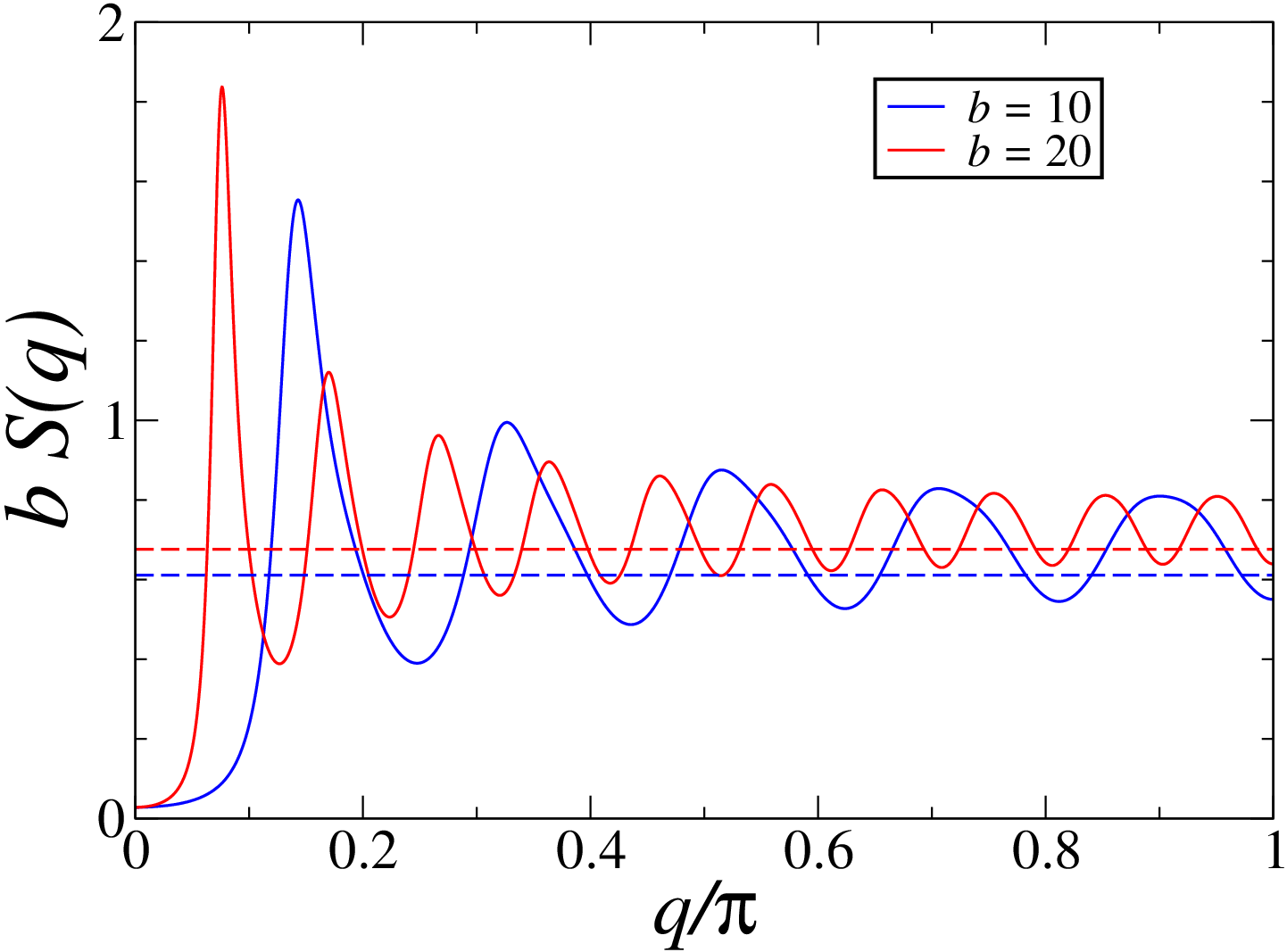}
\caption{\small
Full lines: structure factor $S(q)$ of Rydberg atoms,
multiplied by $b$ and plotted against $q/\pi$,
for $b=10$ (blue) and $b=20$ (red).
Dashed lines with corresponding colors: prediction~(\ref{scst}).}
\label{sosc}
\end{center}
\end{figure}

At variance with the scaling form~(\ref{sksca}),
which falls off as $1/Q^2$ for $Q$ large,
the expression~(\ref{sbsca}) predicts that
\beq
S(q)\approx\rho
\label{scst}
\eeq
is roughly constant and equal to the density of Rydberg atoms
for $\abs{Q}\gg1/b$, i.e., in practice, for $b$ large and $\abs{q}\gg1/b$.
This density is larger than $S(0)$ by a factor growing as $(\ln b)^2$.
The structure factor exhibits oscillations around $\rho$
virtually all over the range of wavevectors,
whose amplitude falls off rather rapidly as the blockade range $b$ is increased
(see figure~\ref{sosc}).
Integrating~(\ref{scst}) over $q$ yields $C_0\approx\rho$,
in agreement with~(\ref{c0c1}) since $\rho$ is small at large $b$.

\section{Discussion}
\label{disc}

This work follows on from a previous paper in collaboration with Krapivsky~\cite{us}.
There, we have put forward an alternative approach,
inspired from the theory of renewal processes,
to study
statistical ensembles of constrained configurations of particles on a one-dimensional lattice,
and especially to determine their configurational entropies.
Here, we demonstrate that a range of observables pertaining to such ensembles,
including chiefly their structure factor and correlation function,
can also be investigated by means of the renewal approach.
It is worth recalling that the scope of this approach is restricted to local constraints
which are expressible in terms of the lengths of clusters of occupied and empty sites.

We have emphasized the pivotal importance of the class of rational models,
which encompasses all examples considered in this paper,
and virtually all situations of physical significance.
Within this rational class,
the complexity of a statistical ensemble is measured
by the degree $\Delta$ of the polynomial $D(z)$ introduced in~(\ref{ncd}).
The integer $\Delta$ is fully analogous to the dimension of the transfer matrix
in the more traditional transfer-matrix approach.
To mention just one example,
the decomposition~(\ref{csum}) of the correlation function
as a sum of $\Delta-1$ decaying exponentials
has a perfect analogue in terms of the subleading eigenvalues of the transfer matrix.

To close, we briefly tackle the situation of ensembles
of blocked configurations beyond the rational class.
To take a striking example, if $\I=\{1,2,4,8,\dots\}$,
i.e., the allowed lengths of particle clusters consist of the powers of two,
the corresponding generating series
\beq
I(z)=\sum_{k=0}^\infty z^{2^k}
\eeq
is a prototypical lacunary series,
having the full unit circle of the complex $z$-plane as a natural boundary.
As a consequence, for a generic set $\J$, the structure factor $S(q)$ has a natural boundary
along some curve in the complex $q$-plane.
The structure of the correlation function
is therefore exceedingly more complex than the finite sum~(\ref{csum}).
It would be desirable to have even one physically motivated model of this kind.

\ack
It is a pleasure to thank Paul Krapivsky for extensive discussions which motivated this work.

\section*{Data availability statement}

Data sharing not applicable to this article as no datasets were generated or analyzed.

\section*{Conflict of interest}

The author declares no conflict of interest.

\section*{Orcid id}

Jean-Marc Luck https://orcid.org/0000-0003-2151-5057

\appendix

\section{The Hendricks-Teller model}
\label{appa}

In this appendix we illustrate
the renewal approach used in the body of this paper on a simpler structural model,
namely the Hendricks-Teller model~\cite{HT}, defined as follows.
Identical pointlike atoms are put along an infinite half-line at random positions~$x_n$
such that $x_0=0$,
and the distances
\beq
\ell_n=x_n-x_{n-1}
\eeq
are independent and identically distributed with some continuous distribution $\rho(\ell)$
such that $\mean{\ell^2}$ is convergent.

The (random) Fourier amplitude $G_N(q)$ and Fourier intensity $S_N(q)$
of the first~$N$ atoms read
\beq
G_N(q)=\sum_{n=1}^N\e^{-\ii qx_n},\qquad
S_N(q)=\frac{\abs{G_N(q)}^2}{N}.
\label{htdefs}
\eeq
Our main interest is in the structure factor
\beq
S(q)=\lim_{N\to\infty}\mean{S_N(q)},
\eeq
which is a self-averaging quantity.

The Fourier amplitudes obey the renewal equation
\beq
G_N(q)=\e^{-\ii q\ell_1}+\e^{-\ii q\ell_1}G_{N-1}(q).
\label{htrenew}
\eeq
The mean Fourier amplitude can be derived
by averaging~(\ref{htrenew}) over the atomic positions.
This yields
\beq
\mean{G_N(q)}=g(q)+g(q)\mean{G_{N-1}(q)},
\eeq
where
\beq
g(q)=\int_0^\infty\e^{-\ii q\ell}\rho(\ell)\,\d\ell
\eeq
is the Fourier transform of the distribution of interatomic distances.
In the limit of an infinitely long sample,
the mean amplitude therefore reads
\beq
\mean{G(q)}=\frac{g(q)}{1-g(q)}.
\label{ght}
\eeq
This result diverges at small wavevectors as
\beq
\mean{G(q)}\approx\frac{1}{\ii q\mean{\ell}}\qquad(q\to0).
\eeq

The structure factor can be derived by multiplying~(\ref{htrenew})
by its complex conjugate, using~(\ref{htdefs}), and averaging over the atomic positions.
This yields
\beq
N\mean{S_N(q)}=1+\mean{G_{N-1}(q)}+\mean{\bar{G_{N-1}}(q)}+(N-1)\mean{S_{N-1}(q)}.
\eeq
In the $N\to\infty$ limit,
the structure factor therefore reads
\beq
S(q)=1+\mean{G(q)}+\mean{\bar{G}(q)}.
\eeq
Using~(\ref{ght}), we thus recover the main result of~\cite{HT} for the present model, i.e.,
\beq
S(q)=\frac{1-g(q)\bar{g}(q)}{(1-g(q))(1-\bar{g}(q))}.
\label{sht}
\eeq

\begin{figure}
\begin{center}
\includegraphics[angle=0,width=0.7\linewidth,clip=true]{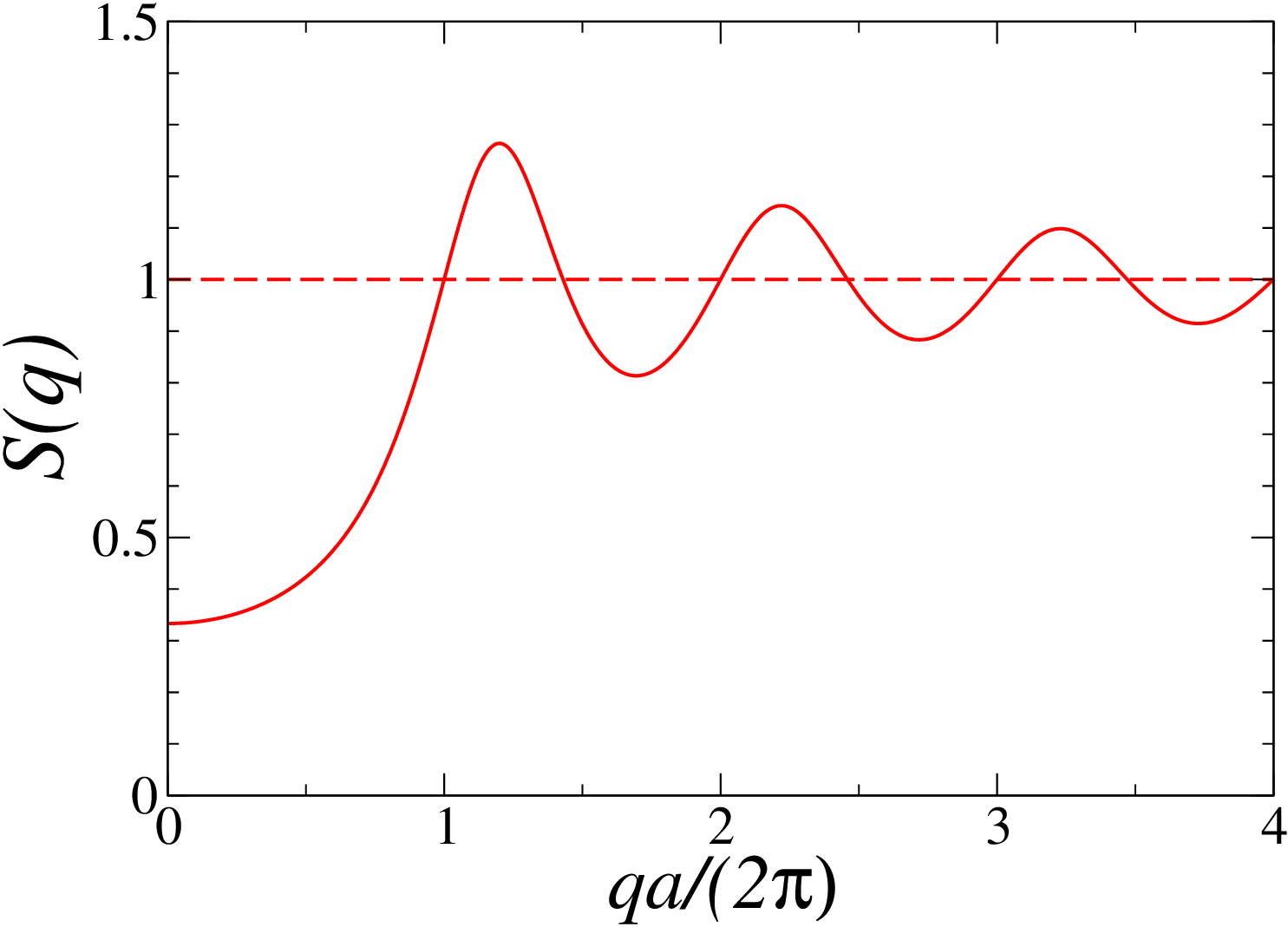}
\caption{\small
Structure factor~(\ref{suni}) of the Hendricks-Teller model
with a uniform distribution of interatomic distances over the interval $[0, a]$,
plotted against $qa/(2\pi)$.}
\label{htuni}
\end{center}
\end{figure}

Let us mention a few properties of the above formula.
Whenever the distance distribution $\rho(\ell)$ is continuous,
its Fourier transform $g(q)$ falls off to zero at large $q$,
and so the structure factor goes to the limit
\beq
S(\infty)=1.
\eeq
Its value at $q=0$,
\beq
S(0)=\frac{\mean{\ell^2}-\mean{\ell}^2}{\mean{\ell}^2},
\label{shtzero}
\eeq
is nothing but the reduced variance of interatomic distances.
This quantity may be either smaller or larger than unity.

The exponential distance distribution $\rho(\ell)=(1/a)\exp(-\ell/a)$
yields a Poissonian distribution of atoms, for which we have $S(q)=1$ for all $q$, as expected.

A simple yet non-trivial example
is provided by a uniform distribution of interatomic distances over some interval $[0, a]$.
The ensuing structure factor
\beq
S(q)=\frac{(qa)^2+2(\cos qa-1)}{(qa)^2+2(1-\cos qa-qa\sin qa)}
\label{suni}
\eeq
is a function of the product $qa$, starting from $S(0)=1/3$
and exhibiting slowly damped oscillations with period $2\pi$ around its limit $S(\infty)=1$
(see figure~\ref{htuni}).

\section*{References}

\bibliography{paper.bib}

\end{document}